E. Konobeevski, A. Kasparov, M. Mordovskoy, S. Zuyev, V. Lebedev, A. Spassky

# Determination of energy of $nn$-singlet virtual state in $d+^2H \rightarrow p+p+n+n$ reaction

**Abstract** The *nn* final state interaction has been investigated in $d+^2H\rightarrow(pp)^S+(nn)^S\rightarrow p+p+n+n$ reaction going through a formation and breakup of singlet diproton and dineutron in the intermediate state. In the kinematically complete experiment performed at $E_d$ =15 MeV two proton and neutron were detected at angles close to those of emission of $(pp)^S$ and $(nn)^S$ systems. Simulation of the reaction showed that the shape of neutron energy (timing) spectrum depends on $\varepsilon_{nn}$ − the energy of virtual $^1S_0$ state of *nn*-system. The most likely value $\varepsilon_{nn} = 0.076 \pm 0.006$ MeV was obtained by fitting procedure using the experimental data and simulations. This low value of $\varepsilon_{nn}$ evidently indicates on effective enhancement of *nn*-interaction in the intermediate state of studied reaction.

### 1. Introduction

Nucleon–nucleon interaction is studied nearly for a century, and a great amount of data on proton–proton (*pp*) and neutron–proton (*np*) interactions has been accumulated over this period. Careful analysis of these data led to constructing NN interaction potentials describing vast majority of experimental data. The situation around neutron–neutron (*nn*) interaction is more ambiguous. In the absence of neutron target, data on this interaction are obtained primarily from reactions with two neutrons in the final state. But in many cases, there are serious discrepancies between available experimental data and the results of the current precise calculations on the base of Faddeev equations.

The most pronounced discrepancies were found in quasifree neutron–neutron scattering. The experimental cross sections obtained at 25 - 26 MeV exceed the theoretical estimates by about 18% [1, 2]. At the same time, the theory describes well the cross sections obtained for quasifree *np* scattering at these energies. Since, at low energies, the quasifree-scattering cross section is dominated by the singlet $^1S_0$ potential component, it was concluded by H.Witala and W Glöckle [3] that this very component was underestimated and this was a reason for the discrepancy between the experiment and theory. An increase of about 8% in this component allows to reach agreement between the theory and experiment for quasifree *np* and *nn* scattering. However, one may notice that authors did not explain the reasons for the introduction of such increase.

According to authors of Dibaryon Model of NN-interaction [4] the strong discrepancies of experiment and theory observed in *nd*- and *pd*-breakup reactions can be explained by a significant strengthening of *nn*- and *pp*-correlations of attractive character in the third nucleon field in $^3H$ (*pnn*) and $^3He$ (*ppn*) systems. New mechanism arising in the Dibaryon model − scalar σ-meson exchange between the nucleon and dibaryon − is presented as a diagram in Fig. 1a. One can propose that such mechanism also may be induced by σ-exchange between two dibaryons ($^1S_0$) in $d+d\rightarrow(pp)^S+(nn)^S\rightarrow p+p+n+n$ reaction (Fig. 1b).

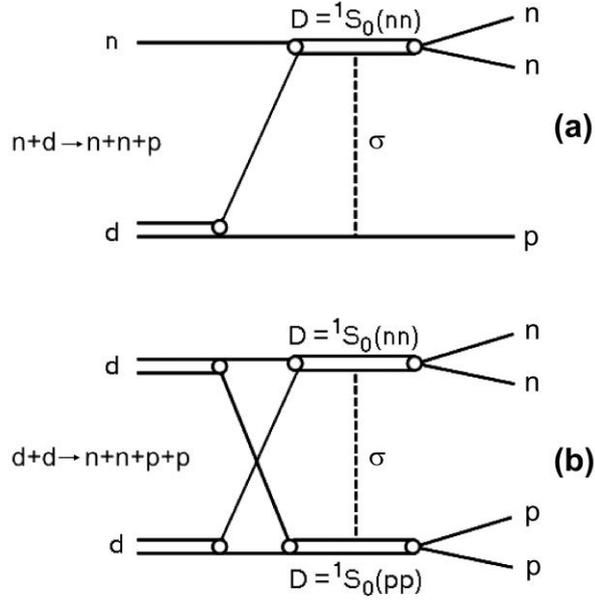

Fig. 1 The graphs illustrating the scalar force induced by σ-exchange between the $^1S_0$ dibaryon and third nucleon (a) and between two $^1S_0$ dibaryons (b).

So we suggest a study of various few body reactions such as $n + {}^3H \rightarrow {}^2H + (nn)$, $n + {}^2H \rightarrow {}^1H + (nn)$, $d + {}^2H \rightarrow (pp) + (nn)$. In these experiments the energy of singlet *nn*- virtual state will be determined in various reactions. Analysis of these energies allows us to estimate the degree of *nn*-correlations in various reactions and determine the mechanism itself of these correlations. As a first part of this investigation we present a study of $d+{}^2H\rightarrow(pp)+(nn)$ reaction, in which *nn*-correlated pair can be produced dynamically in the intermediate state. Thus, measured *nn*-correlation, in particular energy of *nn* virtual singlet state may be different from that for the free *nn*-system.

## 2. Kinematical simulation of $d + {}^2H \rightarrow p + p + n + n$ reaction

The kinematical simulation of the $d + {}^2H \rightarrow (pp)^S + (nn)^S \rightarrow p + p + n + n$ reaction was performed using program intended for studying reactions with three or more particles in the final state [5]. At first we perform simulation of "quasi-binary" reaction $d + {}^2H \rightarrow (pp)^S + (nn)^S$ at deuteron energy of 15 MeV. Thus, masses of two-nucleon systems are taken in a fairly wide range regarding presumed energies of virtual singlet states of *pp*- and *nn*-pair. These calculations allow us to determine the registration angles of the final protons and neutrons. The angles were chosen so that the energy of *pp*-system would be large enough to reduce the ionization losses of final protons, while energy of *nn*-system relatively small in order to reduce the uncertainty in neutron energy determined by time-of-flight technique. It was assumed also that the both protons will be detected by the same detector and the neutron will be detected at the angle close to the angle of emission of *nn*-system.

Further simulation was performed for four-body kinematics at emission angles of neutrons and protons selected at the first stage ($\Theta_{p1,2} = 27° \pm 1.5°$, $\Theta_n = -36° \pm 1.2°$). As a result of simulation, we get an array of events with all parameters of four final particles (energies, momenta, emission angles). Additionally, using the information on the angles and energies of both neutrons we can construct the value of relative energy of two neutrons that is the excess energy of *nn*-system above the threshold of its breakup into two neutrons:

$$\varepsilon_{nn} = \frac{1}{2} \cdot \left( E_1 + E_2 - 2\sqrt{E_1 E_2} \cdot \cos \Delta\Theta \right) \quad (1).$$

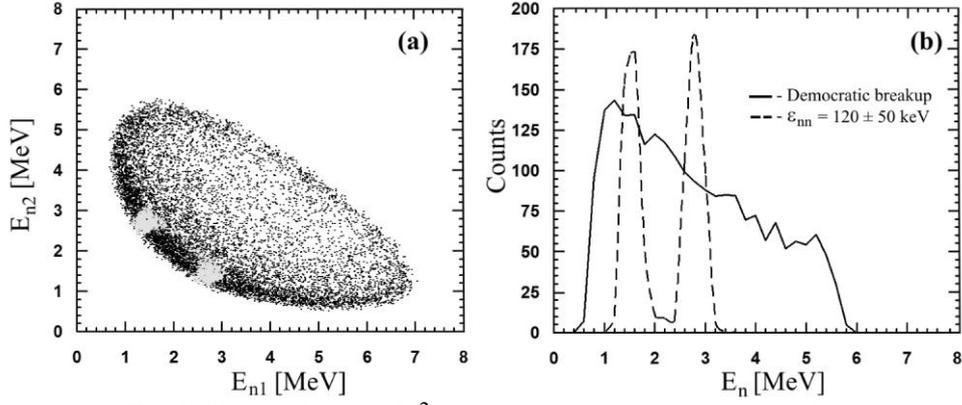

**Fig. 2** Simulation of $d+{}^2H \rightarrow p + p + n + n$ reaction

The data for all values of $\varepsilon_{nn}$ correspond to the so-called democratic breakup. Two-dimensional $E_{n1}$-$E_{n2}$ diagram and neutron energy spectrum for all simulated events are shown in Fig. 2. Selection of events with predetermined values of $\varepsilon_{nn} = E_{nn} \pm \Gamma_{nn}$ leads to structures both in two-dimensional $E_{n1}$-$E_{n2}$ diagram (Fig. 2a) and in neutron energy spectrum (Fig. 2b). The presence of two peaks in this spectrum is due to the fact that in reactions with formation and breakup of NN intermediate state, and under condition that the breakup particle is detected at the angle of emission of NN-system, to hit the detector may only particles emitted in c.m. system or in forward (~ 0°) or in backward (~ 180°) direction. In general, it can be noted that the shape of the energy (and accordingly timing) neutron spectrum is sensitive to the energy of the virtual singlet state, that will allow us to determine this quantity from a comparison of experimental and simulated spectra.

**3. Experiment and results**

The experiment was performed at a 15 MeV deuteron beam of SINP MSU. In the measurement the $CD_2$-target with thickness of 2 mg/cm$^2$ was used. Two protons were detected by a $\Delta E$-$E$ telescope at the angle of 27° while the neutron was detected at -36° (on the other side relative to the deuteron beam) with time-of-flight distance of 0.79 m. These angles correspond to angles of $(pp)^S$ and $(nn)^S$ emission in two body reaction. The two-proton events in the $\Delta E$-$E$ telescope were selected by the ionization losses other than those for protons and deuterons events. At that, in the $\Delta E$-$E$ telescope the total energy of two protons was determined. The neutron energy was determined by the time-of-flight technique.

Thus, from the events corresponding to detection of two protons at 27° and the neutron at -36° the timing spectrum of neutrons was formed. Fig. 3a presents this experimental spectrum in comparison

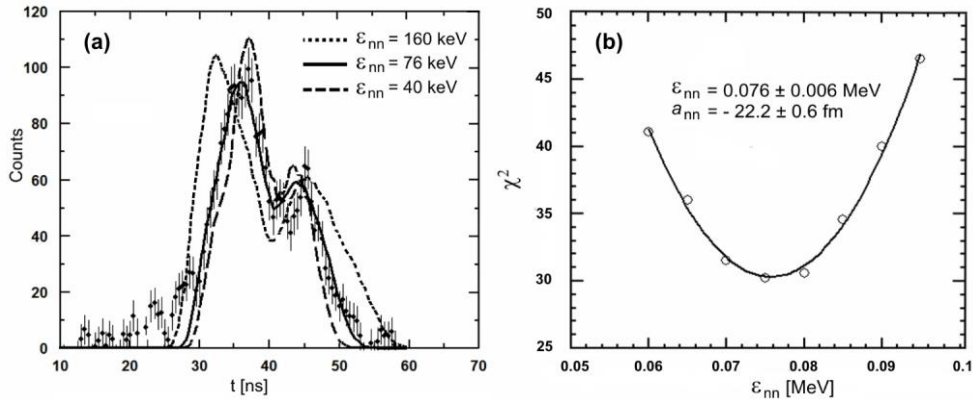

**Fig. 3 a** Experimental and simulated neutron TOF spectra, **b** $\chi^2$ versus $\varepsilon_{nn}$ for the experimental data and simulations with simulated spectra for three values of $\varepsilon_{nn}$ between 0.04 and 0.16 MeV. This figure shows the shape dependence of the neutron timing spectrum as a function of $\varepsilon_{nn}$. For determining the most likely value for $\varepsilon_{nn}$ which is consistent with experimental data we calculated $\chi^2$ using the experimental data and simulations.

Fig. 3b shows a parabolic fit to the $\chi^2$ values obtained at discrete values of $\varepsilon_{nn}$. From this fit a value of $\varepsilon_{nn} = 0.076 \pm 0.006$ MeV was determined, where the quoted uncertainty is based on the $\chi^2_{min} +1$ criterion.

The $\varepsilon_{nn}$ energy is related to the scattering length $a_{nn}$ and the effective range $r_{nn}$ by the equation:

$$\frac{1}{a_{nn}} = -\left(\frac{m_n \varepsilon_{nn}}{\hbar^2}\right)^{1/2} - \frac{1}{2} r_{nn} \frac{m_n \varepsilon_{nn}}{\hbar^2} + ... \qquad (2)$$

According to Eq. (2), the obtained value of the virtual energy $\varepsilon_{nn} = 0.076 \pm 0.006$ MeV corresponds to the value of the singlet $nn$-scattering length $a_{nn}$ = -22.2 ± 0.6 fm. This value of $nn$-scattering length significantly differ from the experimental values of $^1S_0$ nn-scattering length obtained in $nd$-breakup reaction -16.2 ± 0.4 [6], -16.5 ± 0.9 [7], -17.6 ± 0.4 [8], -18.7 ± 0.6 [9], -18.8 ± 0.4 [10].

4. Conclusions

We investigated $d+^2H \rightarrow (pp)^S+(nn)^S \rightarrow p+p+n+n$ breakup reaction, passing through a formation in the intermediate state of dineutron and diproton singlet pairs. For the first time, in a kinematically complete experiment the energy of virtual state of $nn$-system $\varepsilon_{nn}$ is determined. The most likely value $\varepsilon_{nn} = 0.076 \pm 0.006$ MeV was obtained from the $\chi^2$ fit using the experimental data and simulations. The corresponding value of the $nn$-scattering length is much greater than values obtained in $nd$-reaction that evidently indicates an effective enhancement of $nn$-interaction in the intermediate state of studied reaction.

**Acknowledgements**. The authors acknowledge valuable discussions with Prof. V.I. Kukulin. This work was supported in part by RFBR under Grant No 16-32-00743 мол_а.

References
1. Siepe, A., et al.: Neutron-proton and neutron-neutron quasifree scattering in the *n-d* breakup reaction at 26 MeV. Phys. Rev. C 65, 034010 (2002).
2. Ruan, X.S., et al.: Experimental study of neutron-neutron quasifree scattering in the *nd* breakup reaction at 25 MeV. Phys. Rev. C 75, 057001 (2007).
3. Witala, H., Glockle W.: The *nn* quasifree *nd* breakup cross section: Discrepancies with theory and implications for the $^1S_0$ *nn* force. Phys. Rev. C 83, 034004 (2011)
4. Kukulin, V.I., et al.: New mechanism for intermediate-and short-range nucleon-nucleon interaction. J. Phys. G: Nucl. Part. Phys. 27, 1851 (2001)

5. Zuyev, S.V., et al.: A program for simulation experiments to study reactions with three particles in the final state. Bull. Russ. Acad. Sci. 78, 345 (2014)
6. Huhn, V., et al.: New investigation of the neutron-neutron and neutron-proton final-state interaction in the *n-d* breakup reaction. Phys. Rev. C 63, 014003 (2000).
7. von Witsch, W., et al.: Neutron-neutron final-state interaction in the $^2$H (*n, p*) 2*n* reaction at $E$n = 17.4 MeV. Phys. Rev. C 74, 014001 (2006).
8. Crowe, B.J., et al.: Determination of *nn* and *np* Scattering Lengths from Coincidence-Geometry *nd* Breakup Cross-Section Data. TUNL Progress Report. XLV, 65 (2005-06).
9. Gonzales Trotter, D.E., et al.: New Measurement of the $^1S_0$ Neutron-Neutron Scattering Length Using the Neutron-Proton Scattering Length as a Standard. Phys. Rev. Lett. 83, 3788 (1999).
10. Gonzales Trotter, D.E., et al.: Neutron-deuteron breakup experiment at $E$n = 13 MeV: Determination of the $^1S_0$ neutron-neutron scattering length $a_{nn}$. Phys. Rev. C 73, 034001 (2006).